\documentclass[iop]{emulateapj}
\usepackage{graphicx}    
\usepackage{color}

\newcommand{\Msun}{M_{\odot}}

\newcommand{\Mvir}{M_{\rm vir}}
\newcommand{\Vvir}{V_{\rm vir}}
\newcommand{\Rvir}{R_{\rm vir}}

\newcommand{\NHI}{N_{\rm HI}}

\submitted{}

\begin{document}
\title{Angular Momentum Acquisition in Galaxy Halos}
\author{Kyle R. Stewart\altaffilmark{1,2},
        Alyson M. Brooks\altaffilmark{3},
        James S. Bullock\altaffilmark{4,5},
        Ariyeh H. Maller\altaffilmark{6},
        J\"urg Diemand\altaffilmark{7},
        James Wadsley\altaffilmark{8},
        Leonidas A. Moustakas\altaffilmark{2}
        }

\altaffiltext{1}{Department of Natural and Mathematical Sciences, California Baptist University, 8432 Magnolia Ave., Riverside, CA 92504, USA}
\altaffiltext{2}{Jet Propulsion Laboratory, Pasadena, CA 91109, USA}
\altaffiltext{3}{Department of Astronomy, University of Wisconsin--Madison, 475 N. Charter St., Madison, WI 53706}
\altaffiltext{4}{Center for Cosmology, Department of Physics and Astronomy, The University of California at Irvine, Irvine, CA, 92697, USA}
\altaffiltext{5}{Center for Galaxy Evolution, Department of Physics and Astronomy, The University of California at Irvine, Irvine, CA, 92697, USA}
\altaffiltext{6}{Department of Physics, New York City College of Technology, 300 Jay St., Brooklyn, NY 11201, USA}
\altaffiltext{7}{Institute for Theoretical Physics, University of Zurich, 8057, Zurich, Switzerland}
\altaffiltext{8}{Department of Physics and Astronomy, McMaster University, Main Street West, Hamilton L85 4M1, Canada}

\begin{abstract} {
We use high-resolution cosmological hydrodynamic simulations to study the angular momentum acquisition of gaseous halos around
Milky Way sized galaxies. We find that cold mode accreted gas enters
a galaxy halo with $\sim70\%$ more specific angular momentum than dark matter
averaged over cosmic time (though with a very large dispersion).  In fact, we find that all matter has a higher spin parameter when measured at accretion than when averaged over the entire halo
lifetime, and is well characterized by
$\lambda \sim 0.1$, at accretion.  Combined with the fact that cold flow gas spends a relatively short time (1-2 dynamical times) in the halo before sinking to the center, this naturally explains why cold flow halo gas has a specific angular momentum much higher than that of the halo and often forms ``cold flow disks''.  We demonstrate that the higher angular momentum of cold flow gas is related to the fact that it tends to be accreted along filaments.

}
\end{abstract}

\section{Introduction}
\label{Introduction}

In the modern Lambda Cold Dark Matter (LCDM) paradigm of galaxy formation, dark matter halos form via the continuous
accretion of diffuse material, as well as mergers of smaller dark matter halos over time
\citep[e.g.,][and references within]{Stewart08,FakhouriMa09,Diemer12}.
While the acquisition of angular momentum in dissipationless $N$-body simulations
has been well studied  \citep[e.g.,][]{Bullock01,Vitvitska02,Maller02,DOnghia07,Avila-Reese05,Bett07, Bett10},
our understanding of the extent to which gas behaves in the same manner has only recently started
to be explored computationally  \citep{Chen03, SharmaSteinmetz05, Brook10, Kimm11, Pichon11, Sharma12}
In this paper, we extend such work by looking at angular momentum acquisition in galaxy halos in a suite of
cosmological hydrodynamic simulations that allow us to directly measure the angular momentum not only of dark matter, but also of the gaseous halos that are built from ongoing gas accretion from the cosmic web.

In this LCDM framework, the canonical picture of galaxy formation is that gas falling
into a dark matter halo shock-heats to the virial temperature
of the halo \citep{Silk77, WhiteRees78, BarnesEf87, WhiteFrenk91, MallerBullock04}, spending a considerable amount of time in the galactic halo and redistributing angular momentum so that
gas in the halo has the same specific angular momentum as the underlying dark matter.
This shock-heated gas can then cool to the center of the halo
where it presumably forms an angular momentum supported galactic disk.
Under this formalism, extended gaseous halos around galaxies are expected to have similar specific angular momentum as the dark matter
(by construction) and the size of galactic disks should be directly proportional to the spin\footnote{For the sake of brevity, we often use the term ``spin'' throughout this paper as a synonym for specific angular momentum,
except when specifically referring to the ''spin parameter,'' $\lambda$ (see Eq. \ref{eq:spinparameter}).}
of the dark matter halo \citep{FallEfst80,MoMaoWhite98}.
Indeed, this picture of gas accretion and angular momentum acquisition
is a vital component of most semi-analytic models of galaxy formation
in order to model galaxies with realistic disk sizes based on underlying dark matter halo properties
\citep[recently, e.g.,][]{Cattaneo06,Croton06,Somerville08,Dutton12}.
However, abundance matching techniques \cite[e.g.,][]{Behroozi10} have demonstrated that at most $\sim20\%$ of
available baryons can be found in galaxies in the present day, thus the angular momentum of a galaxy may
strongly depend on the origin of the baryons that form the disk \citep{MallerDekel02}. In fact, recent simulations find
that the gas in galactic disks has higher spin than that of the dark
matter \citep{Chen03, Brooks11} in agreement with the model of \cite{MallerDekel02}.

One possible reason for this discrepancy is the treatment of gas accretion.
Recent advances in hydrodynamic simulations as well as analytic arguments have stressed the importance of
``cold-mode'' accretion of gas that does not shock-heat to the virial temperature before building the central galactic
disk.  This mode of accretion occurs when the cooling time of infalling gas is shorter than the compression time,
making it difficult to establish a stable shock for halos masses below a critical value\footnote{``Cold-mode'' accretion is also important
in halos above this critical mass at high redshift, but we will focus on $z<2$ for the majority of this paper.} of $\sim10^{12}\Msun$
\citep[e.g.,][]{Keres05, DekelBirnboim06, Brooks09, Dekel09, FGKeres10, Stewart11a, vandeVoort11, FG11}.
This cold mode of accretion tends to have a very high angular momentum content
\citep{Keres09,KeresHernquist09,Agertz09,Brook10,Pichon11,Tillson12},
with cold-mode gas in galaxy halos typically having $\sim2$--$5$ times more
specific angular momentum than the corresponding dark matter in the halo \citep{Stewart11b,Kimm11},
with a possible exception for cold-mode accretion at extremely high redshifts ($z>6$), which may provide 
preferentially low angular momentum materials to galactic bulges \citep{Dubois12}.

The goal of this paper is to investigate angular momentum acquisition at moderate and low redshifts ($z<3$) in a suite of
fully cosmological high-resolution hydrodynamic simulations, in order to understand the origin of angular momentum in gaseous halos of
galaxies---particularly for cold-mode gas---and
to determine how angular momentum acquisition in these \emph{gaseous} halos differs from the
more thoroughly investigated spin of dark matter halos.
While the build-up of high angular momentum gaseous halos invariably influences
the evolution of the galaxies embedded at the center of each halo,
we do \emph{not} focus on the simulated disk galaxies' angular momentum in this work,
as there is already a rich literature of low redshift studies of galaxy kinematics
using simulations similar to those presented here
\citep[e.g.,][]{Governato08,Brooks09,Brook10,Stinson10,Brooks11,Eris}.

The outline of this paper is as follows.  In section \S\ref{simulations} we discuss the details of the cosmological
hydrodynamic simulations used and the effects that each simulated galaxy's unique merger history
has on the more fundamental underlying trends of angular momentum acquisition.
We analyze the angular momentum acquisition into galaxy halos over cosmic time in
section \S\ref{fresh}, and present implications for observations of extended gaseous disks of inflowing
material in \S\ref{coldflowdisks}.  We discuss our results in \S\ref{discussion}, including implications for implementing
cool gas halos with realistic angular momentum content in semi-analytic models of galaxy formation.
We conclude and summarize our findings in \S\ref{conclusion}.

\section{The Simulations}
\label{simulations}

Each of the four simulations analyzed here uses a separate set of cosmological initial conditions to probe galaxies in
different environments, though all simulations use the same fundamental cosmological parameters,
adopted from the best-fit parameters of the WMAP three-year data release \citep{WMAP3}:
$\Omega_{M} = 0.238$, $\Omega_{\Lambda} = 0.762$, $H_{0}= 73 $km s$^{-1}$ Mpc$^{-1}$, $n_s=0.951$, and $\sigma_8=0.74$.
We focus our analysis on the
single most massive galaxy in each simulation, tracking the evolution of each galaxy halo until $z=0$.
Each of our four galaxies reside in a roughly Milky Way size dark matter halo, and we refer to these halos as ``Halo $1$--$4$''
throughout this paper\footnote{Halo $1$ uses the same initial conditions as the very high resolution
$N$-body simulation Via Lactea II, whose merger history can be publicly downloaded at $\texttt{http://www.ucolick.org/\char`\~diemand/vl/}$.
Halos $1$--$2$ are the same halos analyzed in \cite{Stewart11a, Stewart11b}, while Halos $3$--$4$
have also been analyzed in past works \citep[referred to as ``h277'' and ``h285'' respectively in, e.g.,][]{Brooks11}}.
Halos $1$--$2$ both reach the critical halo mass for shock-heating infalling gas ($\Mvir\sim10^{12} \Msun$) at some
point during each simulation, with $\Mvir\sim1.5\times 10^{12} \Msun$ at $z=0$ for both halos.  In contrast,
Halos $3$--$4$ remain below this critical mass at all times, growing to $\Mvir\sim8\times10^{11} \Msun$ at $z=0$.
We define the virial radius of our halos as the radius within which the
average density is $\Delta_{\rm vir}$ times the mean density of the universe,
iteratively removing unbound particles from within halos by running the
Amiga Halo Finder \citep{AHF} on all output snapshots of our simulations\footnote{This unbinding
decreases the virial radius and virial mass by only $\sim5\%$, compared to including unbound particles.},
where $\Delta_{\rm vir}$ is a function of redshift and defined by \cite{BryanNorman98}.
We note that this is a relatively standard method of defining the virial radius of halos
in dissipationless $N$-body simulations \citep[e.g.,][]{Stewart08}, however there is also
a growing concern that conventional definitions of halo virial radius and virial mass are somewhat
arbitrary and that, ultimately, they may not be physically meaningful \citep[e.g.,][]{AnderhaldenDiemand11,Diemer12}.
As we are primarily concerned with comparing the angular momentum acquisition of galaxy halos via different modes (cold-mode gas,
hot-mode gas, and dark matter accretion), the precise definition of the virial radius used should not
impact our results.

We
generate ``zoom-in'' multi-mass particle grid initial conditions
within a periodic box of $40 (50)$ co-moving Mpc on each side for Halos $1$--$2$ ($3$--$4$), in order to account for large scale tidal torques.
The high-resolution particle masses in the initial conditions are $(m_{\rm dark}, m_{\rm gas}) = (17,3.7)\times10^5 \Msun$
for Halos $1$--$2$, and $(m_{\rm dark}, m_{\rm gas}) = (12,2.1)\times10^5 \Msun$ for Halos $3$--$4$.
The simulations use the code GASOLINE \citep{GASOLINE}, which is a smoothed
particle hydrodynamics (SPH) extension of the pure $N$-body gravity
code PKDGRAV developed by \cite{Stadel01}. The gravitational force
softening is $332$ $(347)$ pc for Halos $1$--$2$ ($3$--$4$), which evolves co-movingly until $z = 9$ and
remains fixed from $z = 9$ to the present. The SPH smoothing length
adapts to always enclose the $32$ nearest gas particles and has a
minimum of $0.05$ times the force softening length.

The code assumes a uniform UV background from QSOs, implemented following
\cite{HaardtMadau96} and F. Haardt (2002, private communication).  It includes star
formation, and Compton and radiative cooling as described in \cite{Katz96}.
The ``blastwave'' supernova feedback model implemented
creates turbulent motions in nearby gas particles, keeping them from cooling and
forming stars; this model is implemented in the code by forcibly turning off cooling
for a short time, as described in \cite{Stinson06}.
The only free parameters in this star formation and feedback model are the
minimum density threshold ($\rho_{\rm min}=0.1$ cm$^{-3}$), star formation efficiency factor
($c^{\star}=0.05$), and fraction of supernova energy that couples with the ISM ($\epsilon_{SN}=0.1$
for Halos $1$--$2$ and $\epsilon_{SN}=0.4$ for Halos $3$--$4$)
which have all been motivated by \cite{Governato07} to produce galaxies with
realistic star formation rates, disk thicknesses, gas turbulence, and Schmidt law
over a range in dynamic masses.
Similar simulations utilizing this feedback scheme have shown great success in producing realistic disk-type galaxies
\citep{Governato08}, matching the mass-metallicity relation \citep{Brooks07}, and matching the
abundance of Damped Lyman $\alpha$ systems at $z=3$ \citep{Pontzen08}.
We
refer the reader to \cite{Governato08} for a more detailed description of the simulation code,
though it is also worth noting that most of the analysis in this paper
focuses on gas that has yet to form stars or be directly affected by
supernova feedback.
As we are focusing on angular momentum acquisition in this paper, we also note that
simulations of this kind have produced galactic disks of realistic sizes
when compared to observations \citep{Brooks11}.  

%

With the resolution implemented here, this feedback model results in minimal galactic winds
($\sim100$ km/s) that mostly affect hot gas, and are more prominent in
halos with $\Mvir\lesssim10^{11}\Msun$ \citep{Shen10}.
Stronger outflows result when using high resolution simulations with higher star formation density
thresholds and modified star formation and supernova efficiency parameters
\citep[e.g.,][]{Governato10, Eris, Governato12, PontzenGovernato12},
but the parameters adopted in this paper do not drive outflows of cool gas.  As a result,
we may more easily focus our analysis on the behavior of cold-mode gas \emph{accretion},
which is a fundamental first step to fully understanding the complex environment of gaseous halos
around galaxies, which are thought to include both accretion and outflowing winds.

In the analysis that follows, we define ``cold-mode'' gas particles based on a
temperature cutoff of $250,000$ K, following the characterization put forth by
\cite{Keres05} and \cite{Keres09}.  Gas that has \emph{never} been
above this temperature (until it reaches $0.1\Rvir$) is defined to be ``cold-mode.''
Any gas that does not meet this criterion is termed ``hot-mode.''  We stop tracking temperatures
at $R=0.1\Rvir$ so that these cold/hot mode definitions are not contaminated by
heating due to supernova feedback.  We do not make any special distinction for
cold-mode gas that is associated with mergers.

We note that smooth particle hydrodynamic (SPH) codes are known to have some numerical issues following
cool clouds in a hot medium \citep[e.g.,][]{Agertz09}, however
most of the general results about cold mode accretion have also been confirmed in grid-based adapted mesh refinement
(AMR) simulations, though many small scale details are different.
We believe the results in this paper in regards to angular momentum are robust to numerical implementation
as we are mostly concerned with the state of gas in the the outer halo where resolving
the multi-phase medium is less of a concern.
In addition, \cite{Pichon11} have performed a similar study using the AMR
code Ramses \citep{RAMSES}, and their results are consistent with ours.

\begin{figure*}[tb!]
 \includegraphics[width=1.0\textwidth]{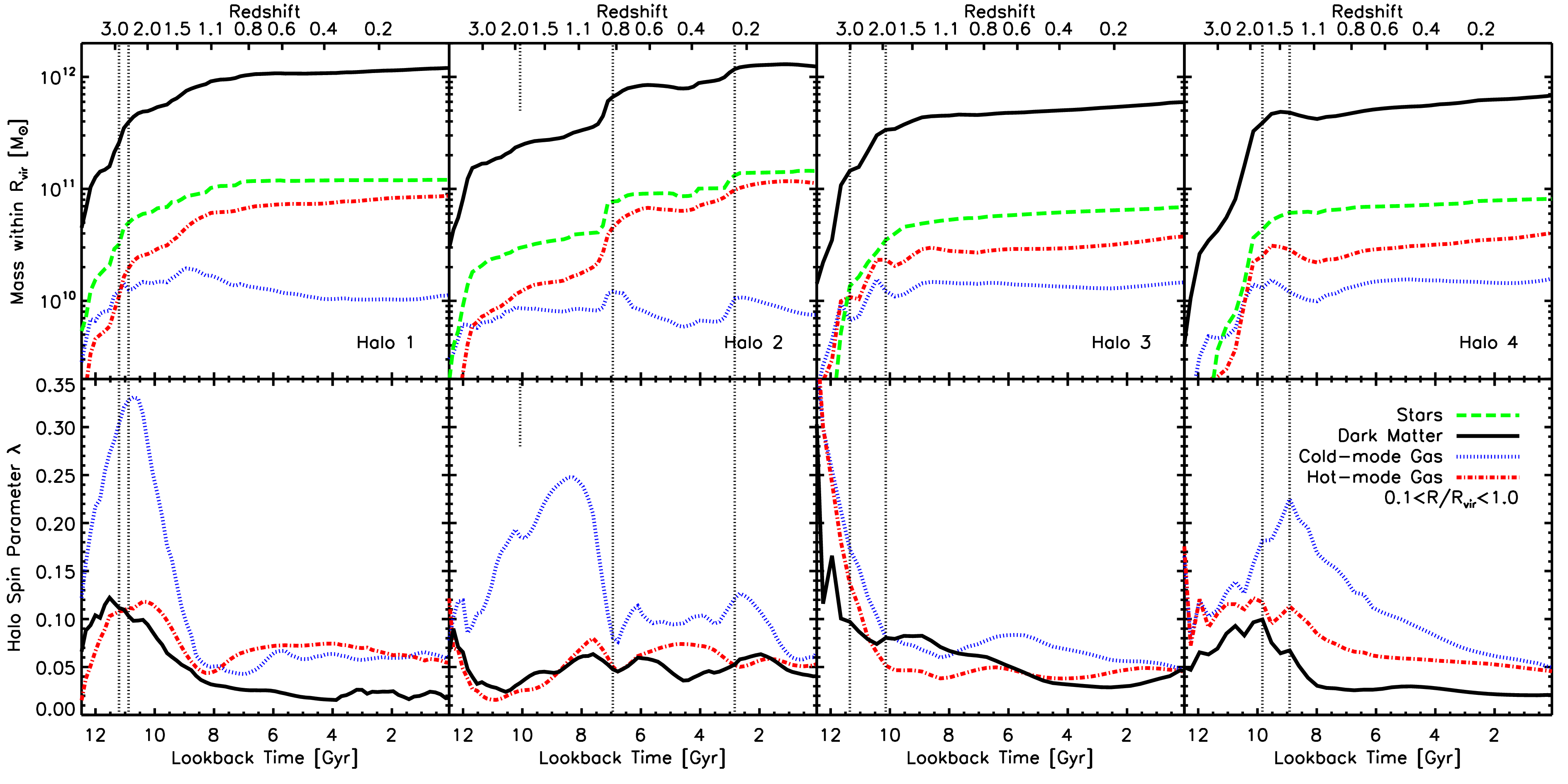}
 \caption{Evolution of mass and angular momentum in the halos of $4$ simulated galaxies over cosmic time
  for cold-mode gas (dotted blue line) hot-mode gas (dot-dashed red)
  stars (dashed green) and dark matter (solid black).
  \emph{Top}: the total mass within the virial radius (after removing unbound particles).
  \emph{Bottom}: specific angular momentum, in terms of the spin parameter, $\lambda$ (see Eq. \ref{eq:spinparameter}).
  Note how drastic changes in angular momentum often correspond to significant merger events (long vertical dashed lines in all panels) or even occasionally by gas-rich minor mergers (short vertical dashed line at $z\sim2$ in Halo $2$).
   }
\label{fig_mass}
\end{figure*}

\subsection{Mergers and Stochasticity}
\label{histories}
The top panels of Figure \ref{fig_mass} show the mass growth as a function of
time for the dark matter as well as the various baryonic components within the halo (all material within the virial
radius, including the galaxy itself).
We note that
there is significantly more cold-mode gas in galaxy halos than there is gas in the galaxies themselves,
for all four simulations and at all epochs.  Thus the cold-mode
gas can largely be thought of as warm halo gas, like that seen in QSO absorption
systems \citep[e.g.,][]{COSHALOS}.

The bottom panels of Figure \ref{fig_mass} show the specific angular momentum of gas and dark matter in each galaxy halo.  Unlike the top
panels, we now remove the galaxy itself from consideration, and only look at material
with $0.1\Rvir<R<\Rvir$.
We quantify the specific angular momentum in this figure in
terms of a dimensionless \emph{spin parameter}, $\lambda$, adopted from \citet{Bullock01}, as:
\begin{equation}
\lambda_x \equiv \frac{j_x}{\sqrt{2} \, V \, R} \,
\label{eq:spinparameter}
\end{equation}
where $\lambda_x$ is the spin parameter of a particular component $x$ (dark matter, hot-mode gas, or cold-mode gas),
$j_x$ is the specific angular momentum ($J/M$) of that component within a sphere of radius $R$,
and $V$ is the circular velocity measured at $R$.  As in past works, we again find that
the spin of cold-mode gas in the halo is typically much higher than that of the dark matter \citep{Stewart11b,Kimm11}.

Throughout this paper, we calculate the specific angular momentum of each component (dark matter, cold-mode gas, hot-mode gas)
in relation to the spin axis of that same component.  In this way, we may more robustly compare the angular momentum
of gas versus dark matter without artificially reducing the spin of one component by projecting it along the axis of a different component.
This approach also makes the most sense observationally, as it is not feasible to project any observed
angular momentum measures of gas in the halos of galaxies with the spin axis of the underlying dark matter, which cannot be observed.
Thus, when the cold-mode and hot-mode gas in our simulations show
similar spin parameters in the bottom left panel of Figure \ref{fig_mass} (for Halo $1$ at $z<1$),
this does not necessarily mandate that each gaseous component
is spinning along the same axis at this time, but that the total \emph{amplitude} of
each component's $J/M$ is similar, regardless of the relative \emph{alignment} with respect to each other
(we will explore the non-trivial relation between the spin axis of these various components in an
upcoming paper).

The tall vertical dotted lines in these figures show when each galaxy experiences a significant merger
event\footnote{A merger of stellar mass ratio greater than  0.3.
These lines correspond to the time of final coalescence, with the understanding that mergers
are not instantaneous events.  It is likely that each merger impacts the halo for several billion years prior to the
times indicated in Figure \ref{fig_mass}.}
and often accompany sharp peaks in the mass growth of the galaxies.  Often, these mergers
significantly re-orient the angular momentum axis of the galaxy, the halo, or both,
indicated by large peaks or dips in the bottom panels of Figure
\ref{fig_mass}.
The short vertical dotted line in Halo $2$ shows a minor gas-rich
merger that seems to have an unusually strong effect on the angular
momentum of the gaseous halo.
It is clear in the bottom panels that the spin of the halos is quite stochastic---especially
for the cold gas in the halo.  In a more detailed sense, these panels show 
that the impact a merger has on the angular momentum of the galaxy halo
depends on the precise characteristics and orbital parameters of the merger.  There are multiple instances
where mergers result in a drastic increase of the specific angular momentum of material in the halo
(for example, Halo $1$ and Halo $4$).  However, there are also times where mergers result in a rapid decrease in the angular
momentum of the halo, due to misalignment between the pre-existing spin axis of the halo and the orbital parameters of the incoming merger
(Halo $2$ at $z\sim0.8$ and Halo $3$).  While the detailed impact that these merger events have
on the angular momentum of the halo would be an interesting topic of future study,
the goal of this paper is to understand the \emph{underlying} steady acquisition
of angular momentum into galaxy halos over cosmic time, apart from the stochastic peaks and dips that result from merger events.
As a result, we will not
be focusing on the detailed minutiae of the angular momentum growth in each of our galaxy halos, which will
depend strongly on the individual accretion history of each halo.

\begin{figure*}[tb!]
 \includegraphics[width=1.0\textwidth]{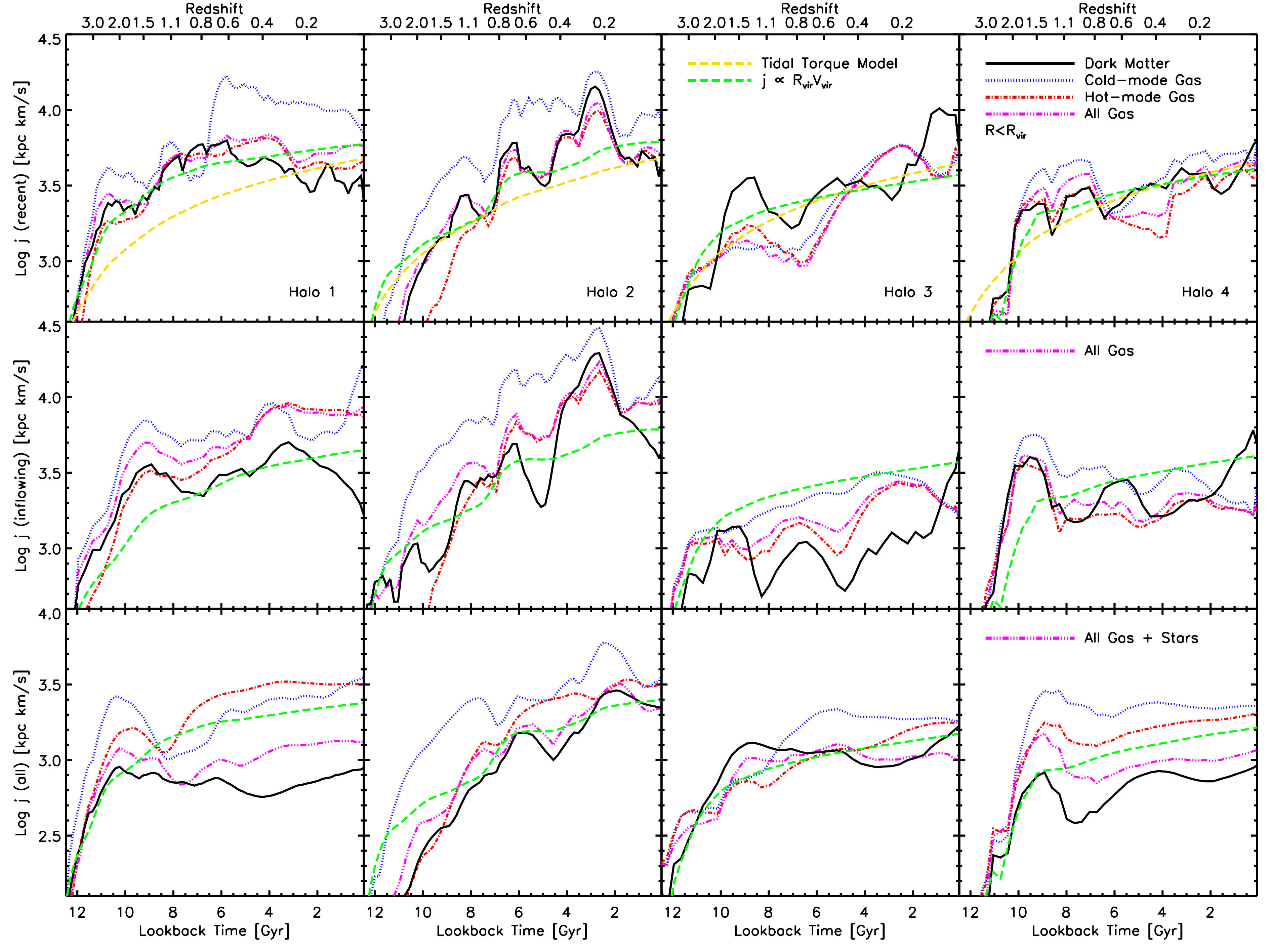}
 \caption{Specific angular momentum of accreted material.
 The dashed green curve represents a semi-analytic model where halo virial quantities are
 known, and a constant spin parameter is assumed, such that
 $j(t) = \lambda \sqrt{2}  \Vvir(t) \Rvir(t)$.  Top panels show only \emph{fresh} accretion---material
 that entered the virial radius within $\sim300$ Myr.  We see that cold-mode gas on average has $\sim50\%$ higher spin than all gas at the time of accretion. We compare our fixed spin parameter
 model ($\lambda=0.1$, green dashed curve) to the more complex tidal torque model of
 \cite{Maller02}, given by the gold dashed curve.  The middle panels
 compute the accretion-weighted average specific angular momentum (see  text for details), again compared to the constant spin parameter model with
 $\lambda=0.1$.  Bottom panels look at \emph{all}   material currently in the halo, compared to the canonical dark
  matter spin model where $\lambda=0.04$.  (Note the difference in scale on the y-axis for the bottom panels.)
   }
\label{fig_dJdt}
\end{figure*}

\section{Angular Momentum Acquisition}
\label{fresh}

We first look at how the angular momentum of fresh gas
accretion to the virial radius compares to the canonical scenario for dark matter.
In the top panels of Figure \ref{fig_dJdt}, we directly calculate the specific angular momentum of recent accretion to the virial radius, for both
gas and dark matter as a function of cosmic time, defining ``recent''
accretion by a timescale of $\lesssim 300$ Myr, which is shorter than the halo
dynamical timescale at all epochs of interest ($z<3$; see \S
\ref{inhalo} for more on gas accretion timescales in the simulations), but long enough
to be practical in terms of storing simulation outputs with a finite timestep resolution.
Particles are considered ``accreted'' onto the halo once they enter
the virial radius of the halo.  On occasion, a particle may enter, exit, and re-enter the halo at a later time; in this
scenario, only the \emph{latest} entry to the halo is considered to be true accretion to the halo,
to avoid double-counting the angular momentum contribution of
any given particle\footnote{At any given time, $<20\%$ of all particles that \emph{ever} entered $\Rvir$ have exited the
virial radius of the system.}.
For comparison, the lower panels of Figure \ref{fig_dJdt} show the total specific angular momentum of \emph{all} material within the
virial radius, including the galaxy itself.

It is well documented that dark matter halos in $N$-body simulations show relatively stable spin parameters of $\lambda\sim0.04$
\citep[recently,][]{Bett07, Maccio07, Berta08, Bett10}.
Since the spin parameter is defined to vary as $\lambda(t) \propto j(t)\Vvir^{-1}\Rvir^{-1}$,
and since we know that for any given halo the circular velocity
($\Vvir$) and the virial radius ($\Rvir$) both grow as a function of cosmic time, the only possible way to maintain a constant
spin parameter is if the specific angular momentum, $j(t)$, increases over cosmic time.
As a result, it is a robust prediction of
LCDM that \emph{the specific angular momentum of freshly accreted material must, on average,
be higher than that of the halo as a whole} \citep[also see][]{Porciani02a,Porciani02b,Maller02,Pichon11}.
This is an important context to keep in mind for much of the following analysis.

\begin{figure}[tb!]
  \includegraphics[width=0.45\textwidth]{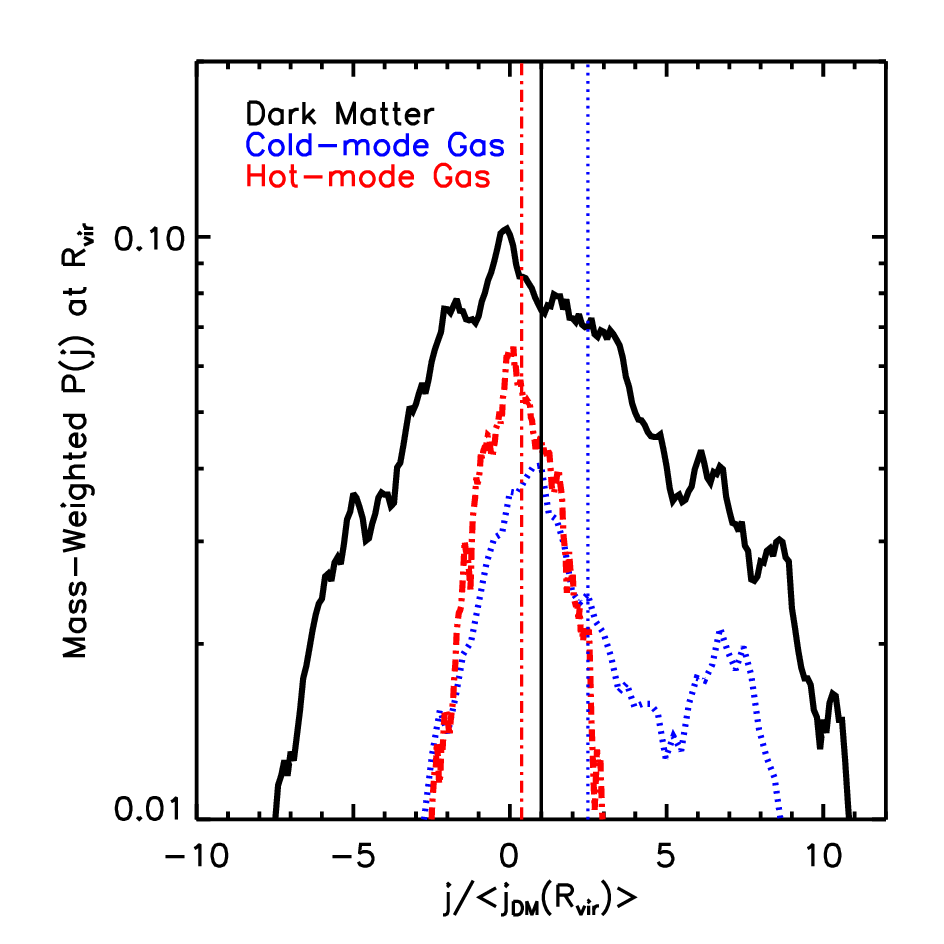}
 \caption{Specific angular momentum distribution, $P(j)$ of dark matter, cold-mode gas, and hot-mode gas at
 $R=\Rvir$ around Halo $2$ at $z\sim1.4$.  To aid in comparison between components,
  the x-axis has been normalized by the average specific angular momentum
  of the dark matter at the virial radius, so that a value of $j/$$<$$j_{\rm DM}(\Rvir)$$>$ $=1$ corresponds to the average
  value for the dark matter curve.
   }
\label{fig_pj}
\end{figure}

\begin{figure*}[tb!]
 \includegraphics[width=1.0\textwidth]{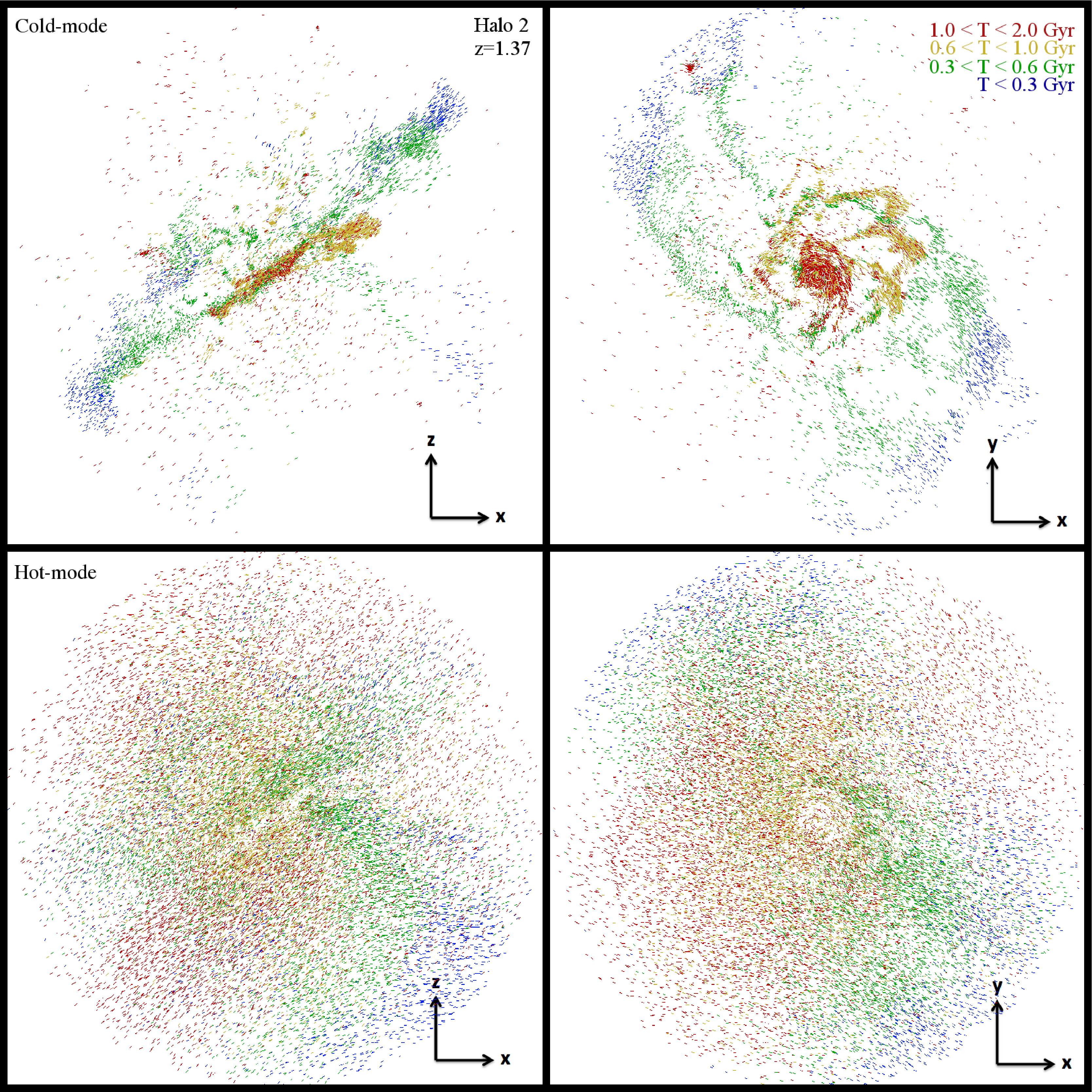}
 \caption{Example velocity flow chart for recent gas accretion.  In each panel, the length of each mark is
  proportional to the projected velocity of the corresponding particle, while the colors correspond to different
  accretion times to the virial radius of the halo.  The width of each panel is $\sim200$ co-moving kpc (the virial
  radius for this halo at this epoch), with left and right panels showing orthogonal projections of the same galaxy.
  The galaxy is viewed edge-on in the left panels and face-on in the right panels, while the top panels show cold-mode accretion
  and the bottom panels show hot-mode accretion.
   }
\label{fig_tipsypic}
\end{figure*}

In each of the top panels, we compare our findings to a simple model where a constant spin parameter of $\lambda=0.1$ is assumed
(i.e., $j\simeq0.14\Rvir\Vvir$) given by the dashed (green) curve,
noting that this is $\sim2.5$ times higher than the global spin parameter of the dark matter.
We also compare to the more complex model (gold dashed curve) of \citet[][]{Maller02}, in which
the angular momentum of fresh accretion, in radial shells around the halo,
has been acquired at turnaround according to linear tidal torque theory \citep[e.g.,][]{Peebles69,Doroshkevich70,White84}.
Remarkably, the simple assumption of $\lambda\sim0.1$  for infalling material produces very similar
curves to the tidal torque model, reproducing the specific angular momentum of fresh accretion
as well as (or better than) the physically motivated tidal torque model.

In the middle panels, we use a complementary approach to probe the spin of accreting material.  Instead of choosing matter with recent infall times, we instead 
compute a velocity-weighted average for $j$ for all particles in the inner halo ($0.5 < R/\Rvir < 1.0$) that posses \emph{in-flowing} radial velocities with 
respect to the halo center: 
\begin{equation}
j({\rm inflowing}) \equiv \left| \frac{ \sum{m_i v_r (\bf{\vec{r_i}} \times \bf{\vec{v_i}}) } }{ \sum{m_i v_r} } \right|, \hspace{1 em}v_r<0
\label{eq:accretionweighted}
\end{equation}
similar to the definition of ``accretion-weighted'' angular momentum performed by \cite{Kimm11}.  While the exact values of $j$ shown in Figure \ref{fig_dJdt} 
depend on the precise definition used (``recent'' versus ``inflowing''), comparison between the top and middle panels 
demonstrates that this quite different and complementary approach for probing inflowing material provides qualitatively similar results.  
We again find that inflowing gas, and particularly cold-mode gas, has significantly more angular momentum than inflowing dark matter in the outer halo of
our simulated galaxies, closer to a spin parameter of $\lambda=0.1$ (shown: green dashed curve) than the canonical value from dark matter simulations of $\lambda=0.04$ (bottom panels).
 
In the bottom panels, the we show the specific angular momentum for \emph{all} particles in the halo, and compare to the 
expectation of $\lambda=0.04$, which is a relatively close approximation to the curves.
The difference between the top/middle versus bottom panels of this figure (a factor of $\sim2.5$ in $j$) is paramount for understanding the
angular momentum content of galaxy halos versus that of galaxies.
We see from this figure that the angular momentum of all baryons in the halo is very similar to that of all dark matter,
both in fresh accretion (top) and in all material within the virial radius (bottom); thus,
the canonical assumption of $\lambda=0.04$ still
seems to be a reasonable estimate for the angular momentum of galaxies, or of all baryons in the \emph{entire} halo.
However, this assumption grossly underestimates the spin of
\emph{recent} (top) or \emph{inflowing} (middle) dark matter accretion into the halo, as well as that of the gaseous component in the halo.
Of particular note is that for each of our simulated galaxies, either in fresh accretion, inflowing accretion, or even
cumulative accretion, cold-mode gas is almost always higher than that of recent dark matter
or even that of hot-mode gas accretion\footnote{A notable exception to this result is
for Halo $3$ at $z\sim1.5$, when $j$ for cold-mode gas accretion is \emph{less}
than that of the recent dark matter accretion, presumably as a result of the mergers
in this systems shown in Figure \ref{fig_mass}.  While some authors distinguish between cold-mode
gas accretion and cold gas associated with merger, we remind the reader that we make no such distinction here.}.
Averaging over all four simulations for the redshift range shown, we find (based on the top panels of Figure \ref{fig_dJdt}) that recent cold-mode gas 
accretion has on average a $\sim70\%$ higher spin than that of
recent dark matter accretion, though with very large variations.
This systematic offset between the angular momentum of cold-mode versus hot-mode gas accretion suggests a 
fundamental kinematic difference between gas that is accreted in each of these modes.

To gain further insight into how these modes of gas accretion differ from each other, and from dark matter accretion,
Figure \ref{fig_pj} shows a representative
mass-weighted histogram of the specific angular momentum for particles of dark matter (solid black), cold-mode gas (dotted blue), and hot-mode gas (dot-dash red),
focusing on a single simulation at a single point in time (Halo $2$ at $z\sim1.4$).
For the sake of straightforward comparison, we normalize the $x$-axis by the average specific angular momentum
of all dark matter at $\Rvir$, $<$$j_{\rm DM}(\Rvir)$$>$\footnote{Under this normalization, a value of $j/$$<$$j_{\rm DM}(\Rvir)$$>$ $=1$ corresponds to the mean value of the
 specific momentum of the dark matter at the virial radius (i.e. the black curve in Figure \ref{fig_pj}) by definition.}, and the $y$-axis by the cosmic baryon fraction.  
We show the distribution $P(j)$ for particles between $0.95\Rvir < R < 1.05 \Rvir$.
The vertical line shows the mean value of $P(j)$.  We see that the specific
angular momentum distribution of dark matter is much broader then that of the gas
components.  The hot-mode gas is much narrower and has a mean near zero, which we might expect if the hot gas is shock heating as it enters the virial radius.  What is more surprising is that the cold-mode gas has a distribution very different than the dark matter with a much higher mean, even though we do not expect the cold-mode gas to have exchanged angular momentum in any coherent fashion.  This dual-peak distribution of cold-mode gas at this epoch is \emph{not} due to a gas-rich merger or some other unusual accretion event, but seems to be a typical result of coherent cold-mode gas accretion along cosmic filaments.  We have checked that these differences remain at $2\Rvir$ suggesting that they have to do with the large scale distribution of the various
components and not particle interactions.
Taken alongside Figure \ref{fig_dJdt}, this suggests a fundamental kinematic difference between the origin of cold-mode gas versus hot-mode gas accretion onto galaxy halos.



We investigate this distinction further by visualizing all recently accreted gas particles within the virial radius
of this same halo at this same epoch in Figure \ref{fig_tipsypic}.
Each panel shows the position of all cold-mode (top) and
hot-mode (bottom) gas particles that were accreted to the halo within the past $2$ Gyr, with the length of each
mark proportional to the projected velocity of each particle, so that the figure gives a qualitative visual impression of
the velocity flow of the particles shown.  The color of each particle corresponds to the accretion time, allowing
a sense of time progression from the most recent accretion (blue; $T<300$ Myr) to the older accretion
that has had time to sink to the center of the halo and mix with the pre-existing gaseous halo (red; $1<T<2$ Gyr).
The galactic disk is edge-on at the center of the image in the left panels, and face-on in the right panels,
as indicated by the red particles in the top plots where cold-mode gas has had sufficient time to reach the galaxy.
In the bottom panels, the galaxy cannot be seen, as hot-mode gas accretion typically takes longer than $2$ Gyr to
reach the galaxy at this epoch (we discuss accretion timescales further in \S\ref{inhalo}).

We note several important qualitative features in Figure \ref{fig_tipsypic}.  The most striking is the stark difference between
cold-mode and hot-mode accretion.  As has been noted in past simulation studies, cold-mode accretion typically flows into the
halo along the direction of the large-scale filaments that fuel the galaxy, most clearly visible in the upper-left panel, while
hot-mode accretion is more likely to be more isotropic \citep[e.g.,][]{Keres09}.  The velocity vectors of the particles
also show that while both cold-mode and hot-mode gas accretion has significant angular momentum on a particle-by-particle
basis, the cold-mode gas in the upper panels shows clear indication of coherent rotation.
While no such bulk motions are readily visible \emph{by simple visual inspection} in the lower panels,
Figure \ref{fig_dJdt} has already demonstrated that the hot-mode gas is also spinning about the galaxy with significant specific angular
momentum---if not quite as high as is present in cold-mode material.

The upper panels also show clear segregation of cold-mode gas particles as a function of accretion time.
As gas first accretes onto the halo, it clearly begins in the outskirts near the virial radius (blue) and gradually falls inward,
spinning about the galaxy and losing orbital energy as it sinks towards the center of the halo (green-gold).
After $1$--$2$ Gyr, the cold-mode gas reaches the galactic
region and begins to fuel the outer edges of the galaxy itself (red).
This clear picture of gas accretion from the cosmic web directly onto the galactic
disk is not indicated by the bottom panels for hot-mode accretion.  Instead, we see significant radial mixing, with gas from a wide
range of accretion times at various radii.

\begin{figure}[tb!]
 \vspace{-1 em}
 \hspace{-3 em}
  \includegraphics[width=0.54\textwidth]{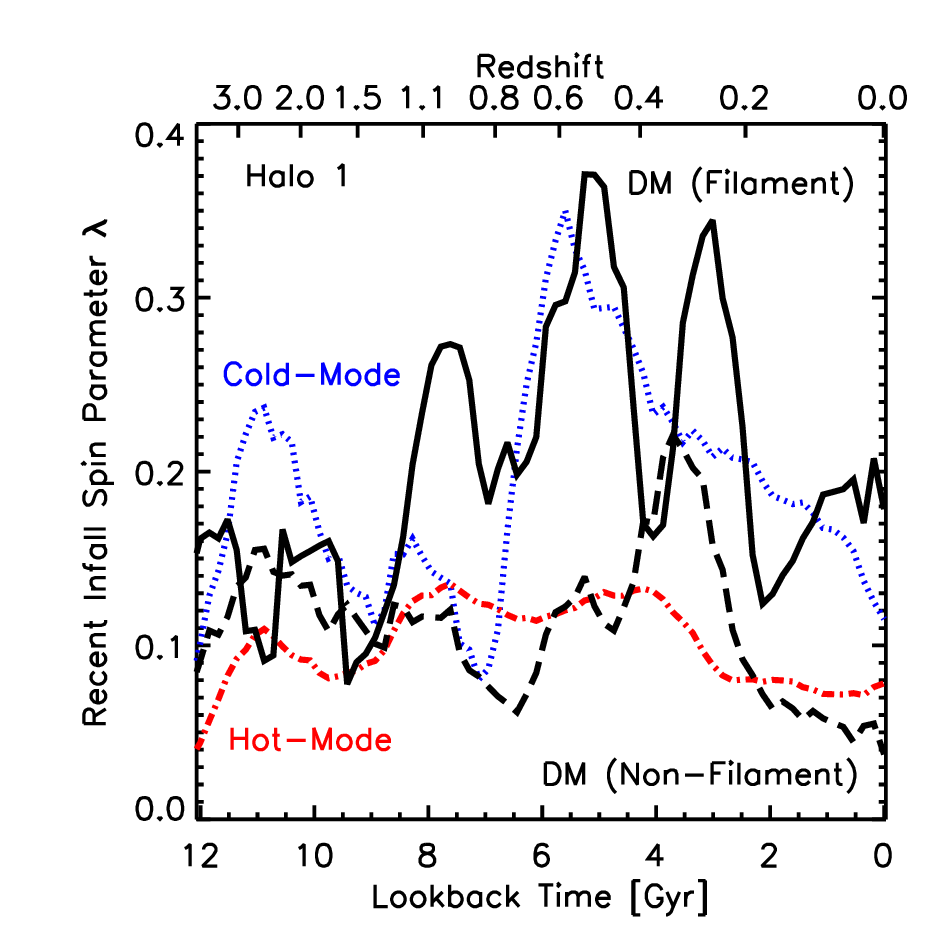}
 \caption{Spin parameter of gas and dark matter, revisited.
 The colored lines show the spin parameter of recent accretion ($T<0.3$ Gyr, $R<0.5\Rvir$).
 When only recent dark matter accretion along cosmic filaments is included, it has a much higher
 angular momentum than the fiducial value of $\lambda\sim0.04$, and is  more
 similar to the spin of recent cold-mode gas accretion, which also flows into the halo along filaments.
 }
\label{fig_spin3}
\end{figure}
If the different spatial and kinematic origins of filamentary (cold-mode) accretion versus
non-filamentary (hot-mode) accretion is truly what leads to the distinct angular momentum distributions on infall,
then we should see a similar distinction between the spin of
filamentary versus non-filamentary \emph{dark matter} accretion as well.
Figure \ref{fig_spin3} shows the spin parameter as a function of time for recently accreted ($T<300$ Myr)
dark matter, cold-mode gas, and hot-mode gas, using an identical color/line scheme as in Figure \ref{fig_mass}.
However, we have now broken up the recent dark matter accretion into two mutually exclusive components:
\emph{filamentary} dark matter accretion\footnote{At each epoch, we
define filamentary accretion spatially by locating particles in a bi-conical region, extending from the halo center
towards the direction of the filament (determined by visual inspection of matter density contours) in either direction,
with a $3$D opening angle of $20$ degrees.} (black solid curve)
versus \emph{non-filamentary} dark matter accretion (black dashed curve).
Aside from some noticeable stochasticity, we see remarkable correlation 
between the specific angular momentum of fresh filamentary accretion (filamentary dark matter versus cold-mode gas) and
between that of fresh non-filamentary accretion (non-filamentary dark matter versus hot-mode gas).
These correlations are highly indicative that the higher specific angular momentum content of fresh cold-mode
accretion is a fundamental consequence of filamentary accretion to dark matter halos
\citep[for a theoretical explanation of \emph{why} filamentary accretion is
expected to have higher angular momentum than non-filamentary
accretion, see][]{Pichon11}.

\begin{figure*}[tb!]
 \vspace{-1 em}
 \hspace{-3.5 em}
 \includegraphics[width=1.05\textwidth]{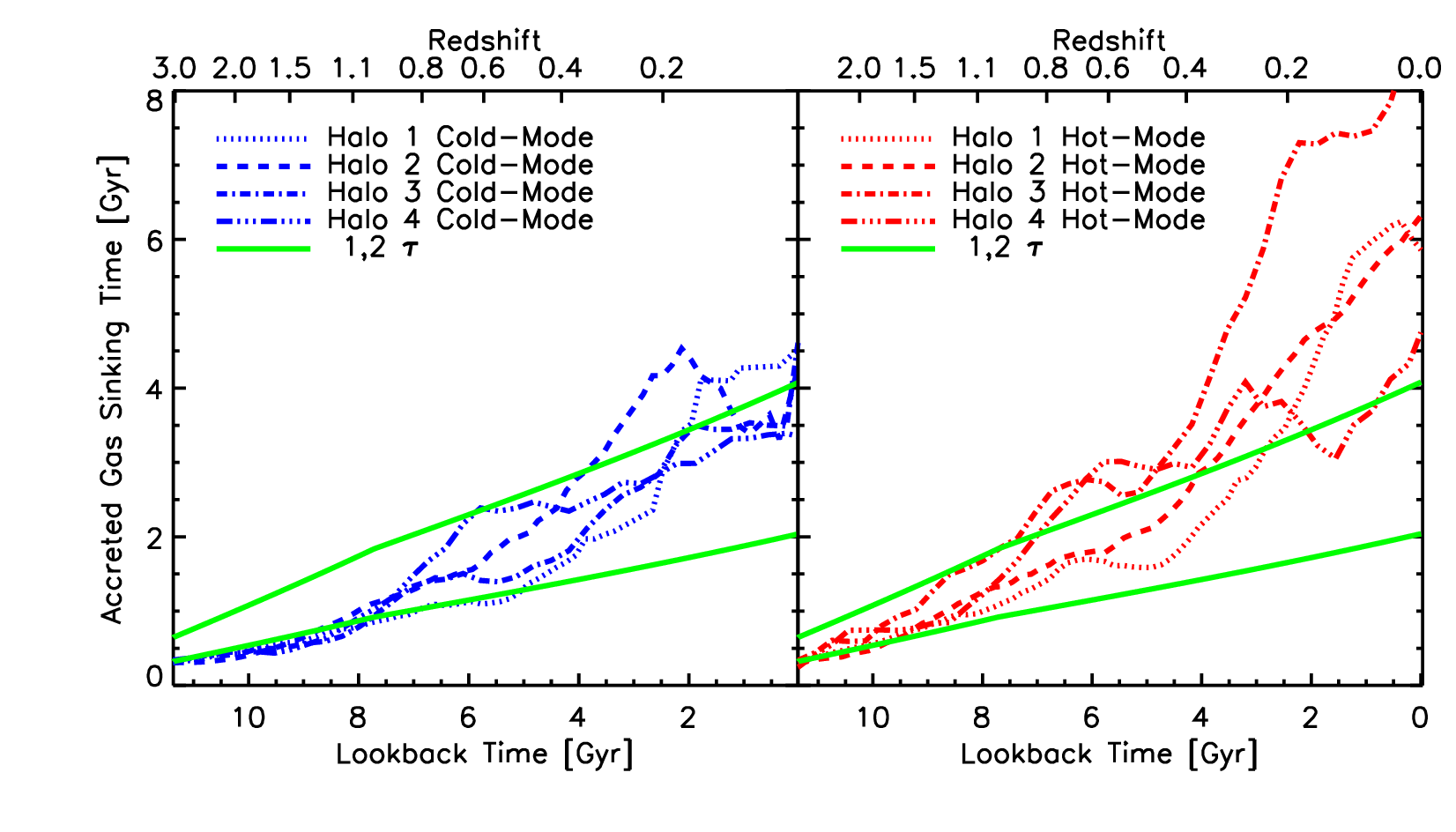}
 \caption{The average amount of time it took for gas currently in the galactic region ($R<0.1\Rvir$)
 to sink from $\Rvir$ to $0.1\Rvir$.
 \emph{Left:} Note that the sinking time for cold-mode gas accretion is relatively short, with timescales typically between $1-2\tau$,
 where $\tau$ is the halo dynamical times and is given by the two solid green curves in both panels.
 \emph{Right:} The sinking time for hot-mode gas accretion is typically $\sim1-2$ times longer than it is for cold-mode gas.
  The median lookback time to accretion for dark matter particles in the halo is considerably longer ($\sim7-10$ Gyr at $z=0$).}
\label{fig_sinktime}
\end{figure*}

Because filamentary accretion is often thought to necessarily indicate highly radial infall trajectories,
we note that this kinematic difference between hot-mode and cold-mode gas is not
simply due to their \emph{average} orbital parameters upon infall.
Quantifying the ``typical'' orbit of fresh accretion to the virial radius in terms of a median \emph{infall angle}---the angle
between a completely radial infall vector and the actual velocity vector of each particle upon first entering the virial radius---we
find that cold-mode gas, hot-mode gas, and dark matter all have comparable mean infall angles ranging from
$\sim30$--$50$ degrees from radial, with no clear trends for any component being consistently more or
less radial than other components.
We want to take a moment to emphasize this point, as it has important ramifications for absorption-line studies of gaseous halos.
When interpreting velocity offsets in studies of absorption systems along lines of sight to background galaxies, it is often assumed
that infalling gas accretion onto the galaxy should always be indicated by high-velocity pure radial infall.
We emphasize that this is not the case; for both dark matter and gas, the vast
majority of accreted material enters the virial radius with high angular momentum, spinning about the halo at an
initial angle of $\sim40$ degrees from a purely radial trajectory, with only a portion of the infall velocity
along the radial direction.
As a result, we advise that absorption-line searches for cold-mode accretion should not focus exclusively on signs of purely radial infall,
but should also look for signs of coherent rotation in cool gaseous halos
\citep[for observable signatures of coherent rotation in Halos $1$--$2$, see][]{Stewart11b}.

\subsection{Halo Gas Accretion Timescales}
\label{inhalo}
Figure \ref{fig_tipsypic} indicated that the radius of infalling cold-mode material is tightly
coupled with its accretion time, whereas the picture is more complicated for hot-mode gas (or dark matter accretion).  In order to understand the processes involved in transporting high angular momentum
fresh accretion from the virial radius to the galaxy, it is important that we recognize the different
timescales involved in this transport.
However, 
we first must ask the question:
once infalling material falls to a given radius, does it \emph{stay} within that
radius?  If material routinely enters the galactic region only to return along an elongated orbit to the
outer edges of the halo, it will be difficult to define an ``infall timescale'' in a meaningful way.

We compare the cumulative mass of all material
that was \emph{ever} accreted to within
a given radius, $R$, to the mass \emph{currently} enclosed within $R$, and
find that the vast majority of gas and dark matter that falls within the virial radius is ultimately bound to the halo,
with $\lesssim20\%$ of all mass
that ever entered the halo being lost to the system, most likely during violent major mergers.
In contrast, when we focus on material that has fallen within the inner halo ($R<0.5\Rvir$) we notice an important difference.
While $\sim85$--$95\%$ of all gas that has ever entered the inner halo remains there, only $\sim60$--$70\%$
of dark matter is equally bound to this inner radius.

The difference between this behavior of dark matter and gas within the halo is not surprising, as the
highly elliptical elongated orbits of collisionless dark matter particles will often bring them
back to the outer halo for significant periods of time during each orbit about the halo's center.
In contrast, high angular momentum gas particles are more likely to mix with other gas in the inner halo, redistributing
angular momentum so that they no longer have enough orbital energy to exit
the inner halo. 
Consequently, while it is indeed straightforward to define a ``sinking time'' for
infalling gas,
it is not meaningful to define a similar quantity for dark matter.
Thus, for each halo we define the sinking time
by tracking every gas particle \emph{currently} in the galactic region of the halo at each epoch ($R<0.1\Rvir$) back in time
in the simulations, calculating the amount of time it took to reach $R=0.1\Rvir$ after first entering the virial radius.

\begin{figure*}[t!]
 \hspace{1 em}
 \includegraphics[width=1.0\textwidth]{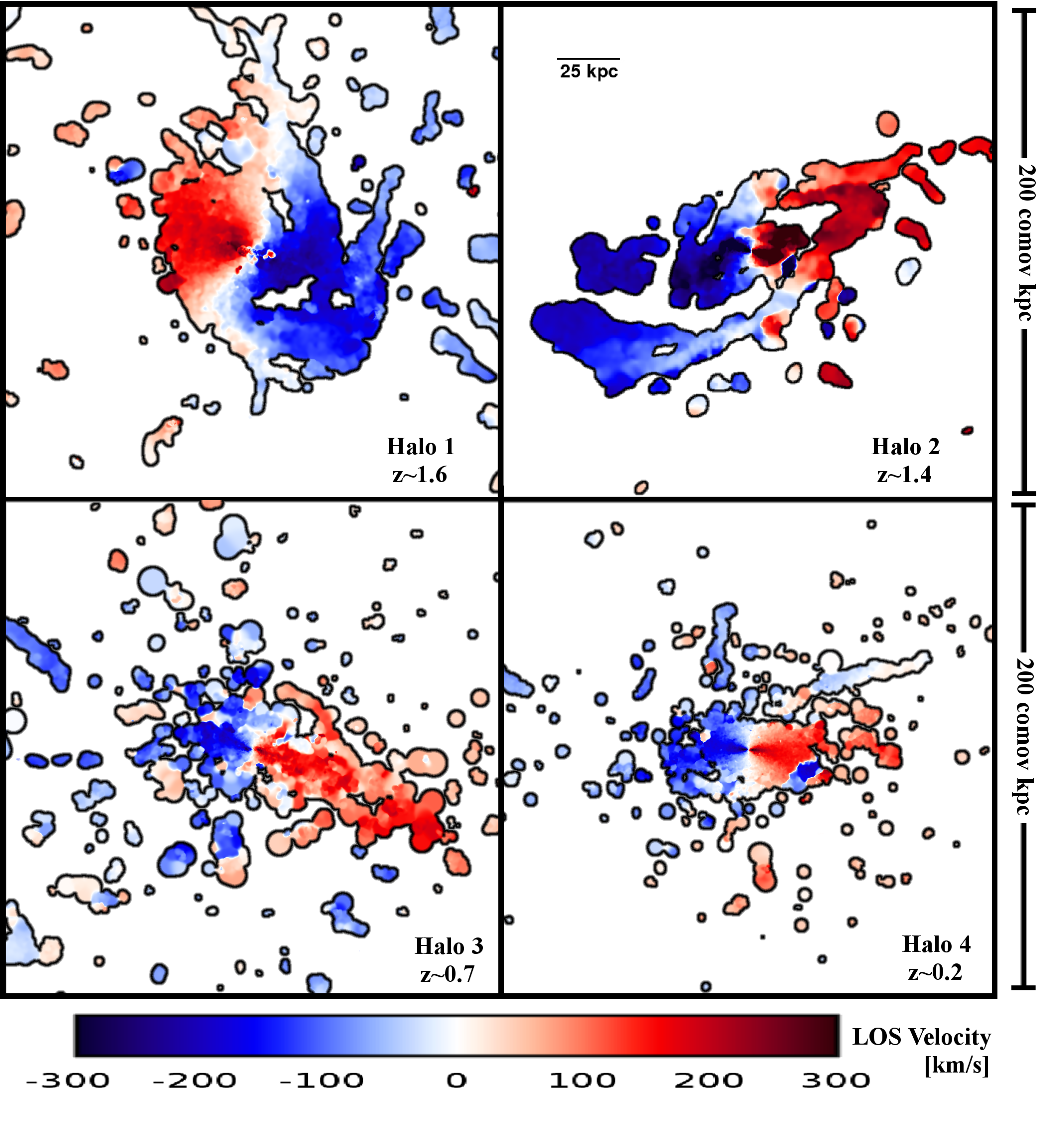}
 \caption{``Cold flow disks'' in each of our four simulated halos, at $z\sim1.6, 1.4, 0.7, 0.2$ respectively
 going from the top left to the bottom right panel.  All panels are on the same velocity and co-moving distance scale, with
 the width of each panel set to $200$ co-moving kpc.
 Note the coherent rotation in the cool halo gas well beyond the inner regions where the galaxy resides.
 The galactic disk is viewed near edge-on in each panel (\emph{i}$>65$).
 }
\label{fig_coldflowdisks}
\end{figure*}

Using this definition, Figure \ref{fig_sinktime} shows the mass-weighted
mean sinking time for cold-mode gas particles in each of our simulations as a function of time.
Because the sinking timescale invariably depends on halo properties, which in turn depend on the
epoch in question, we find it useful to compare these timescales to the halo dynamical time\footnote{We
define the halo dynamical time by the crossing time: $\tau = R/V \propto (\Delta_v(z) \, \rho_u(z))^{-1/2}$, such
that $\tau$ is independent of halo mass and only depends on redshift. We adopt simplified fitting
functions for $\tau(z)$ as presented in \cite{Stewart09a}.  At redshifts $z=3,2,1,0$, $\tau(z)\simeq0.3,0.5,1,2$ Gyr respectively.}
$\tau(z)$.
We compare the sinking time to $\tau$ and $2\tau$ (the solid green curves), 
finding that typical sinking times are relatively short for cold-mode gas---only
$\sim1$--$2\tau$ ($<3$ Gyr at $z=0$). 
We repeat this analysis to determine the average sinking time from $R=1.0$--$0.5\Rvir$, for all material currently
in the inner halo ($R<0.5\Rvir$).  
Not surprisingly, the sinking time for material still in the inner halo is roughly half of the values
shown in Figure \ref{fig_sinktime}.
These findings strongly suggest that infalling cold-mode gas
in galaxy halos is not a cumulative reservoir of ancient past accretion, but instead
probes \emph{fresh} accretion to the halo, with very high specific angular momentum
content similar to that presented in the top panels of Figure \ref{fig_dJdt}.

The sinking time for hot-mode gas accretion is typically longer than that of the cold-mode curves shown
here by a factor of $\sim1-2$.
As mentioned previously, we cannot provide a true ``sinking time'' comparison for dark matter
particles, but we instead calculate the median look-back time
to first accretion for dark matter particles currently in the halo at $z=0$ and find
considerably longer timescales: $\sim7-10$ Gyr for dark matter particles in $0.1\Rvir<R<1.0\Rvir$
and $\sim5-8$ Gyr for dark matter particles in the outer halo ($0.5\Rvir<R<1.0\Rvir$).

\section{Observational Consequences}
\label{coldflowdisks}
Based on our findings, we know that fresh cold-mode gas accretion onto galaxy halos has much more specific angular momentum
than is typical for the entire dark matter halo (Figure \ref{fig_dJdt}, top), or for all baryons in galaxies (Figure \ref{fig_dJdt}, bottom),
and that fresh cold-mode gas accretion even has higher specific angular momentum than recent hot-mode gas accretion,
due to its fundamentally different origin along cosmic filaments
(Figures \ref{fig_dJdt}--\ref{fig_spin3}).
It is possible that these inherent difference in the angular momentum content of cold-mode
versus hot-mode accretion might play a role in explaining why spiral and elliptical galaxies appear to retain
different percentages of the specific angular momentum of all infalling material \citep{RomanowskyFall12}.

In addition, we find that infalling gas does not travel through
the halo in purely radial orbits, but rather has typical infall angles of $\sim40$ degrees (with respect to radial), with
the spin of cold-mode gas in the halo often in a coherent direction, set by the direction of the large-scale cosmic filaments
(Figures \ref{fig_tipsypic}-\ref{fig_spin3}).  We have also established that cold-mode gas in galaxy halos
has relatively short sinking timescales, and thus probes fresh accretion (Figure \ref{fig_sinktime}).

Taking these results together, it is not surprising that cool halo gas around galaxies often
forms coherent structures with
very high angular momentum, formed from continually infalling material from the cosmic web that eventually fuels the galaxy.
As a result, we find that cold-mode gas accretion often forms extended ``cold flow disks'' of dense gas that
rotates coherently about the galaxy, in roughly the same direction as the galaxy.
Previous simulation studies have demonstrated that
this gas accretion should be observable as extended, warped disks of cool gas dense enough to form neutral hydrogen
\citep[e.g.,][]{Roskar10,Stewart11b}, which may help explain
observations of XUV disks \citep{Thilker05,Thilker07} and extended HI disks \citep[][]{GarciaRuiz02,Oosterloo07, Walter08}
around galaxies, which tend to co-rotate with the inner galactic disks
\citep{ChristleinZaritsky08}.
Even in less extreme cases, observations by \cite{Wang13} find that for a
sample of roughly Milky Way size 
galaxies that have been selected to be HI-rich, there is often evidence of 
extended, clumpy HI disks that suggest an orderly accretion
process involving a range of high angular momentum gas---qualitatively
very similar to what we have seen in our simulations.

While cold flow disks are not universal phenomena at all times in our simulations, neither are they rare isolated incidents.
Figure \ref{fig_coldflowdisks} gives a visual impression of
the rotation of cool gas within $R<100$ co-moving kpc over a broad redshift range and sampling each of our four simulations,
instead of only focusing on a single halo at a single epoch as we did in Figure \ref{fig_tipsypic}. 
For each pixel in the images, the
blue--red shading corresponds to the mass-weighted average velocity of HI gas along the line of sight, where we only
show sightlines above a minimum detectability threshold of $\NHI>10^{16}$ cm$^{-1}$.
All four galaxies show clear signs of coherent rotation of gas in the halo, as well as
co-rotation between the gaseous halo and the galactic disk---even out to large radii.
The co-rotation of the galaxies and their cold flow disks are a natural consequence of the
high angular momentum content of cold-mode gas in the halo, coupled with the fact that cold-mode gas accretion
quickly sinks to the center of the halo, feeds the galactic disk, and thus helps define the
angular momentum direction of the galaxy over time.

Based on our simulations, we predict that cold flow disks (possibly already observed as XUV disks or extended HI disks)
should be more common for galaxies below the critical
shock-heating threshold $\Mvir\lesssim10^{12}\Msun$, though they will likely extend farther into the halo
for galaxies very close to this threshold \citep{Stewart11a}. They should contain more specific angular momentum than their associated
galaxies or dark matter halos, and are likely to be more metal-poor than the associated galaxies
(since the material is fresh accretion from the cosmic web).  In addition, since cold flow disks probe recent cold-mode
gas accretion to the halo, the orientation of these disks (often misaligned with the central galaxy) may give better
indication of the direction of large scale cosmic filaments in which halos are embedded than the orientations of
galaxies themselves.  We will expand on this conjecture further in an upcoming paper.


\section{Discussion}
\label{discussion}
In the canonical picture of angular momentum acquisition of galaxies, it is assumed that the angular momentum of baryons (galaxies)
are set by the underlying spin of the dark matter halo in which each galaxy resides \citep[e.g.,][]{MoMaoWhite98}.  This formalism
has led to great successes in reproducing galaxy population using semi-analytic models of galaxy formation in which galaxies
and their properties are added on top of underlying $N$-body simulations of dark matter structure \citep[e.g.,][]{Somerville08}.

The assumptions inherent in this canonical picture of
angular momentum acquisition \emph{do not work for gaseous halos} of galaxies.
While galaxies represent a cumulative growth of material over long time scales (similar to the underlying dark matter),
gaseous halos probe the most recent accretion of fresh material from the cosmic web---especially
for cool gas resulting from cold-mode accretion, which has very short sinking times of only a few billion years.
To further complicate matters, cold-mode and hot-mode accretion contribute different amounts of specific angular momentum,
with cold-mode accretion having higher specific angular momentum upon infall due to its distinct filamentary nature.

Based on the results presented here, we propose a new scenario for angular momentum acquisition, in which: $1)$ cold-mode gas
typically carries more specific angular momentum than hot-mode gas accretion---a fundamental difference that appears to be linked
to its filamentary accretion nature $2)$ the sinking time of cold-mode gas is quite short, comparable to the halo dynamical time
$3)$ fresh accretion from the cosmic web
(and thus, gas present in the halos of galaxies) has a much higher spin than that of the galaxy or the dark matter halo,
spinning about the halo as it falls in, often in the form of extended cold flow disks
$4)$ it is this high angular momentum gas that provides galaxies with both mass and angular momentum, helping to set the
spin axis of the galaxy.

We note that in detail the full picture of angular momentum acquisition will be more complex than the
scenario outlined above, in part due to the presence of galaxy scale gaseous outflows,
which have been emphasized in recent observations of cool halo gas
\citep[e.g.,][]{Steidel96, Martin05, Rubin10}
and were not included in the simulations presented in this paper.
It is worth noting, however, that a recent simulation by \cite{Brook10}---one that includes bi-conical
galaxy-scale outflows as a result of their high resolution and modified feedback scheme---found that the high angular momentum inflowing
cold-mode gas was \emph{least} affected by the outflowing gas.  This was primarily due to the fact that
inflowing material was more likely to approach the galaxy along its major axis, while outflowing material
was more likely to contain low angular momentum gas
ejected along the minor axis of the galaxy.
Thus, the two mechanisms---cold-mode inflow and ejected outflows---typically
inhabit different spatial regimes within the halo.

\subsection{Semi-analytic Models of Galaxy Formation}
\label{SAM}
While recent semi-analytic models have begun implementing cold-mode and hot-mode gas accretion into galaxies
\citep[e.g.,][]{Croton06,Cattaneo06,Somerville08},
most of these models still presume that the angular momentum of galaxies are primarily
set by the spin parameter of the dark matter within the halo.
Though this may provide a reasonable approach for estimating stellar disk sizes (because these are the baryons that
were accreted early enough to have already formed stars), it will be difficult (if not impossible) to explain the presence
of massive extended HI disks around galaxies with typical halo spin parameters \citep{GarciaRuiz02,Oosterloo07, Walter08}
or to accurately model the velocity structure of cool gas halos in the circum-galactic regions of galaxies
without accounting for the fact that recent accretion is spinning more quickly than the bulk of the dark matter in the halo.

We propose that semi-analytic models should consider estimating the angular momentum of cool gas halos
relatively simply by adopting the higher spin parameter (some $\sim3$ times higher than the canonical spin of the halo)
for all recently accreted material, motivated by the top panels of Figure \ref{fig_dJdt}.

Since this cold-mode halo gas is continually accreting onto the galaxy itself, one could also model
the angular momentum of fresh gas accretion onto the outer edges of the galaxy (that is, cold-mode gas that is currently falling
on to the galaxy for the first time) by an angular momentum $j(t)\sim \lambda \sqrt{2} \Vvir(t-\tau)\Rvir(t-\tau)$, where $\lambda\sim0.1$,
and $\tau$ is the sinking time (Figure \ref{fig_sinktime}), comparable to the halo dynamical time.
However, we note that it is non-trivial to integrate this continuous fresh accretion (to the galactic region)
to obtain the total angular momentum of the resulting galaxy.
Over time, this high-spin material experiences significant vector cancelation, due to the change in relative
orientations between the galaxy, the dark matter, the cold-mode halo gas, and the large-scale cosmic
filaments from which new material is constantly being accreted \citep[see][]{Kimm11}.
These orientations do not stay fixed over cosmic time, but vary somewhat stochastically based on large
scale tidal fields and non-linear torques during major mergers.  Instead, the direction between the galaxy orientation, the entire halo,
and that of the inflowing cosmic filaments are more likely to be offset by $\sim20-40$ degrees, instead of being perfectly aligned
\citep[e.g.,][in preparation]{vandenBosch02,SharmaSteinmetz05,Kimm11}.

\section{Conclusion}
\label{conclusion}
In this paper, we used four high resolution cosmological hydrodynamic simulations to investigate the acquisition of angular momentum in
roughly Milky Way sized galaxy halos over time.  Motivated by recent findings that the specific angular momentum of gaseous halos are
considerably higher than that of the dark matter halo \citep{Stewart11b,Kimm11} we focus on how reservoirs of cold-mode gas in galaxy halos
grow and evolve over time, and how the angular momentum of this material differs from the canonical scenario for angular momentum growth in
galaxies.  We note that the angular momentum history of galaxy halos is sometimes quite stochastic---especially
during galaxy mergers that may reorient the angular momentum axis of the system---so we focus on the average behavior of each
system over time. 
We summarize our findings below.

\begin{enumerate}

   \item Cold-mode gas enters the halo preferentially along the direction of cosmic filaments, while hot-mode accretion is more isotropic
    in origin.  As a result, cold-mode gas inherently has more specific angular momentum at first infall than hot-mode gas accretion ($\sim 70\%$ more).
    When filamentary dark matter accretion is compared to non-filamentary accretion, we find similar differences between the initial
    specific angular momentum of dark matter upon infall.

  \item The spin of all gas within the halo (including the galaxy) is broadly consistent with well-studied dark matter simulations, which find
    a relatively stable value of $\lambda\sim0.04$ over a broad range in epoch and mass.  However, when we focus only on \emph{fresh}
    accretion---gas and dark matter that has only been in the halo for $<300$ Myr---we find much higher spins, consistent with
    $\lambda\sim0.1$.

  \item While gas accretion onto galaxy halos is often pictured as purely radial infall, we find that this is universally \emph{not}
    the case in our simulations.  Due to the high angular momentum of infalling material, it always spins about the halo
    with a typical infall angles (with respect to radial) of $\sim30-50$ degrees at the virial radius.


  \item While the elongated orbits of collisionless dark matter particles may allow dark matter to populate the halo
    (even at the outskirts, near the virial radius)
    for long periods of time, inflowing gas is relatively short-lived in the halo.  The ``sinking time'' for cold-mode gas to travel
    from the virial radius to accretion to the galactic region ($R<0.1\Rvir$) is $\sim1$--$2$ halo dynamical times ($<3$ Gyr at $z=0$).

  \item This high angular momentum, quickly sinking, filamentary cold-mode accretion often takes the form of extended cold flow disks
    of newly accreted inflowing material, coherently rotating roughly in the same direction of the galaxy.  These extended gaseous disks
    may be related to observations of XUV and extended HI disks around galaxies.
\end{enumerate}

The nature of gas cooling in simulations invariably depends on the exact simulation code used,
with different codes producing slightly different results from one another depending on the
implementation and resolution, as well as different schemes for modeling the effects of supernova feedback,
and in what way it injects energy back into the ISM.  As a result, the definition of
what gas is considered cold-mode versus hot-mode could be argued to be a somewhat arbitrary distinction,
especially when defined by a single maximum temperature cut, as we have done in this paper.

Nevertheless, the very nature of our results suggest a fundamental underlying difference between the accretion modes we
have labeled ``cold-mode'' and ``hot-mode,'' as the two accretion mechanisms show distinct properties
(even before first infall to the halo virial radius) in terms of
spatial position and angular momentum content upon entering the halo.
Whether or not the cold streams break up on their way to the disk, the fact that they enter with high
angular momentum is likely a robust physical distinction.  Based on the results presented here,
we speculate that it may in fact be more physically meaningful to consider the bi-modal mechanism for
gas accretion onto galaxies not as fundamentally ``cold'' versus ``hot,'' (though this certainly appears to be true),
but rather as ``filamentary'' versus ``non-filamentary'' modes of gas accretion onto galaxy halos.
We admit, however, that the robustness of any such reclassification would require a much more thorough
analysis of filamentary versus non-filamentary accretion than what we have presented here,
and would be a useful topic of further study.


\acknowledgements
Halos $1$ and $2$ were run on the Cosmos computer cluster at JPL, and the Greenplanet computer cluster at UC Irvine.
Resources supporting this work were provided by the NASA High-End Computing (HEC) Program through
the NASA Advanced Supercomputing (NAS) Division at Ames Research Center.
KRS was partially supported by an appointment to the NASA Postdoctoral Program at the Jet Propulsion Laboratory,
administered by Oak Ridge Associated Universities through a contract with NASA.
This research was partially carried out at the Jet Propulsion Laboratory, California Institute of Technology,
under a contract with the National Aeronautics and Space Administration.
AB acknowledges support from The Grainger Foundation.
JSB was partially supported by NASA grant NNX09AG01G.
JD has been supported by the Swiss National Science Foundation (SNF).
LAM acknowledges NASA ATP support.
Copyright 2013. All rights reserved.

\bibliography{angmom}

\end{document}